\documentclass[final]{aipproc}

\layoutstyle{6x9}

\begin{document}


\title{Onset of Convection on a Pre-Runaway White Dwarf}

\author{L.~J.~Dursi} { 
   address = {Center for Astrophysical Thermonuclear Flashes, The University of Chicago, Chicago, IL 60637} 
   ,altaddress = {Department of Astronomy and Astrophysics, The University of Chicago, Chicago, IL 60637}
}
\author{A.~C.~Calder}{
   address = {Center for Astrophysical Thermonuclear Flashes, The University of Chicago, Chicago, IL 60637}
   ,altaddress = {Department of Astronomy and Astrophysics, The University of Chicago, Chicago, IL 60637}
}
\author{A.~Alexakis}{
   address = {Center for Astrophysical Thermonuclear Flashes, The University of Chicago, Chicago, IL 60637}
   ,altaddress = {Department of Astronomy and Astrophysics, The University of Chicago, Chicago, IL 60637}
}
\author{J.~W.~Truran}{
   address = {Center for Astrophysical Thermonuclear Flashes, The University of Chicago, Chicago, IL 60637}
   ,altaddress ={Department of Astronomy and Astrophysics, The University of Chicago, Chicago, IL 60637}
}
\author{R.~Rosner}{
   address={UMBC/GEST Center, NASA/GSFC, Greenbelt, MD 20771}
   ,altaddress = {Center for Astrophysical Thermonuclear Flashes, The University of Chicago, Chicago, IL 60637}
}
\author{M.~Zingale}{
   address = {Department of Astronomy and Astrophysics, The University of California, Santa Cruz, CA 95064}
   ,altaddress = {Center for Astrophysical Thermonuclear Flashes, The University of Chicago, Chicago, IL 60637}
}
\author{B.~Fryxell}{
  address = {Enrico Fermi Institute, The University of Chicago, Chicago, IL 60637}
}
\author{P.~M.~Ricker}{
   address = {Center for Astrophysical Thermonuclear Flashes, The University of Chicago, Chicago, IL 60637}
   ,altaddress = {Department of Astronomy and Astrophysics, The University of Chicago, Chicago, IL 60637}
}
\author{F.~X.~Timmes}{
   address = {Center for Astrophysical Thermonuclear Flashes, The University of Chicago, Chicago, IL 60637}
   ,altaddress ={Department of Astronomy and Astrophysics, The University of Chicago, Chicago, IL 60637}
}
\author{K.~Olson}{
   address={UMBC/GEST Center, NASA/GSFC, Greenbelt, MD 20771}
   ,altaddress = {Center for Astrophysical Thermonuclear Flashes, The University of Chicago, Chicago, IL 60637}
}

\begin{abstract}
Observed novae abundances and explosion energies estimated from
observations indicate that there must be significant mixing of the heavier
material of the white dwarf (C+O) into the lighter accreted material
(H+He). Accordingly, nova models must incorporate a mechanism that will
dredge up the heavier white dwarf material, and fluid motions from an
early convection phase is one proposed mechanism.

We present results from two-dimensional simulations of classical nova
precursor models that demonstrate the beginning of a convective phase
during the `simmering' of a Nova precursor.   We use a new hydrostatic
equilibrium hydrodynamics module recently developed for the adaptive-mesh
code FLASH.  The two-dimensional models are based on the one-dimensional
models of Ami Glasner\cite{Glasner1997}, and were evolved with FLASH
from a pre-convective state to the onset of convection.  \end{abstract}
\maketitle 

\section{Introduction}

As a classical nova precursor accretes material from its neighbor, it
heats up; by the time its peak temperature becomes roughly $4 \times 10^7
{\mathrm ~K}$ -- well before the final stages of runaway -- the accreted
atmosphere becomes convectively unstable.  The resulting convective
motions may be important for the process of dredging up white dwarf
material into the accreted atmosphere.

In this paper, we examine the turn-on of convective motions in a white
dwarf atmosphere based on one-dimensional early-time models provided
to us by Ami Glasner.  This initial model is the same used in other
multidimensional studies \cite{Glasner1997,Kercek1998}, but taken at an
earlier time -- at the last timestep before the onset of convection in
the 1-d model code.  We map this model into the multidimensional FLASH
code \cite{flash} using techniques developed in \cite{hse}, and
perturb the models to investigate the onset of convective motions.

\section{Simulations}

\begin{figure}[tbp]
  \includegraphics[angle=90,width=\textwidth]{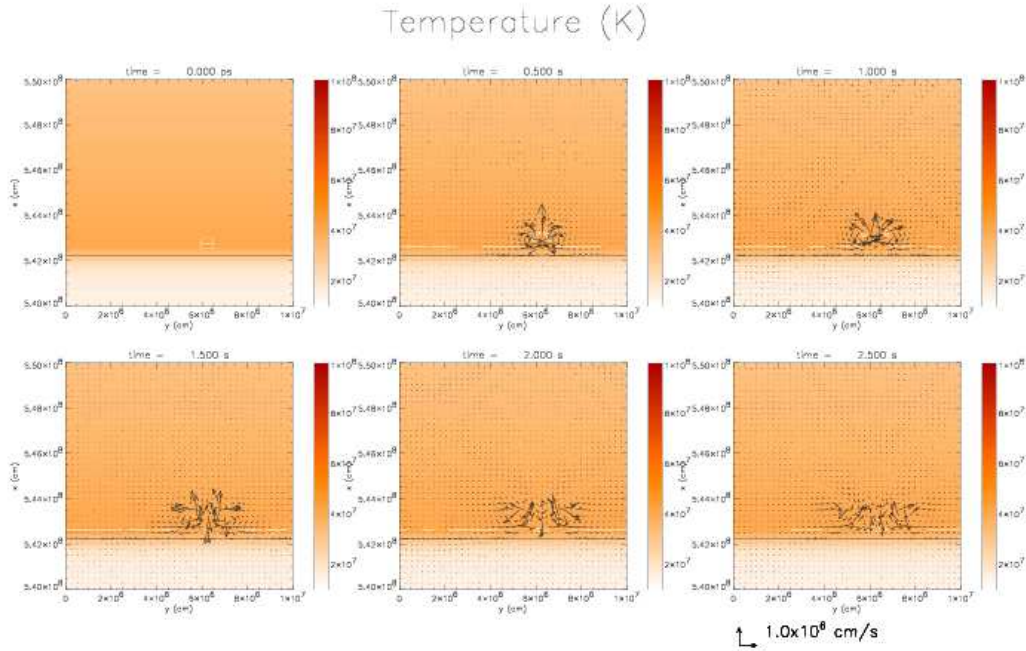}
  \caption{The 2D evolution of the first convective roll, seeded by a
           5\% temperature perturbation at the hottest point in the
           atmosphere.  The white contour surrounds the `hot spot' --
           $ T > 4.01 \times 10^7 {\mathrm {K}}$.  Sound-wave transients
           from the initial perturbation are visible.   Typical induced
           velocities are $v \sim 10^6 {\mathrm {cm~s^{-1}}}$ in the
           rolls, $v \sim 10^5 {\mathrm {cm~s^{-1}}}$ at the interface.}
  \label{fig:5percenttimeseries} 
\end{figure}

Figure \ref{fig:5percenttimeseries} represents early convective motions
in the atmosphere.  The temperature in a $ 5~\mathrm{km} \times 5
\mathrm{km}$ region at the hottest (and least convectively unstable)
point in the atmosphere is initially increased by 5\%.  Sound waves
are emitted, and a convective roll begins.  This is shown in Figure
\ref{fig:5percenttimeseries}, through approximately one rollover time,
at times $t = 0,.2,.4,.6,.8, \mathrm{and} 1.0~\mathrm{s}$ after the
perturbation is imposed.  Note that the velocity field extends to the
C+O interface.

\begin{figure}[ht]
  \includegraphics[width=0.32\textwidth]{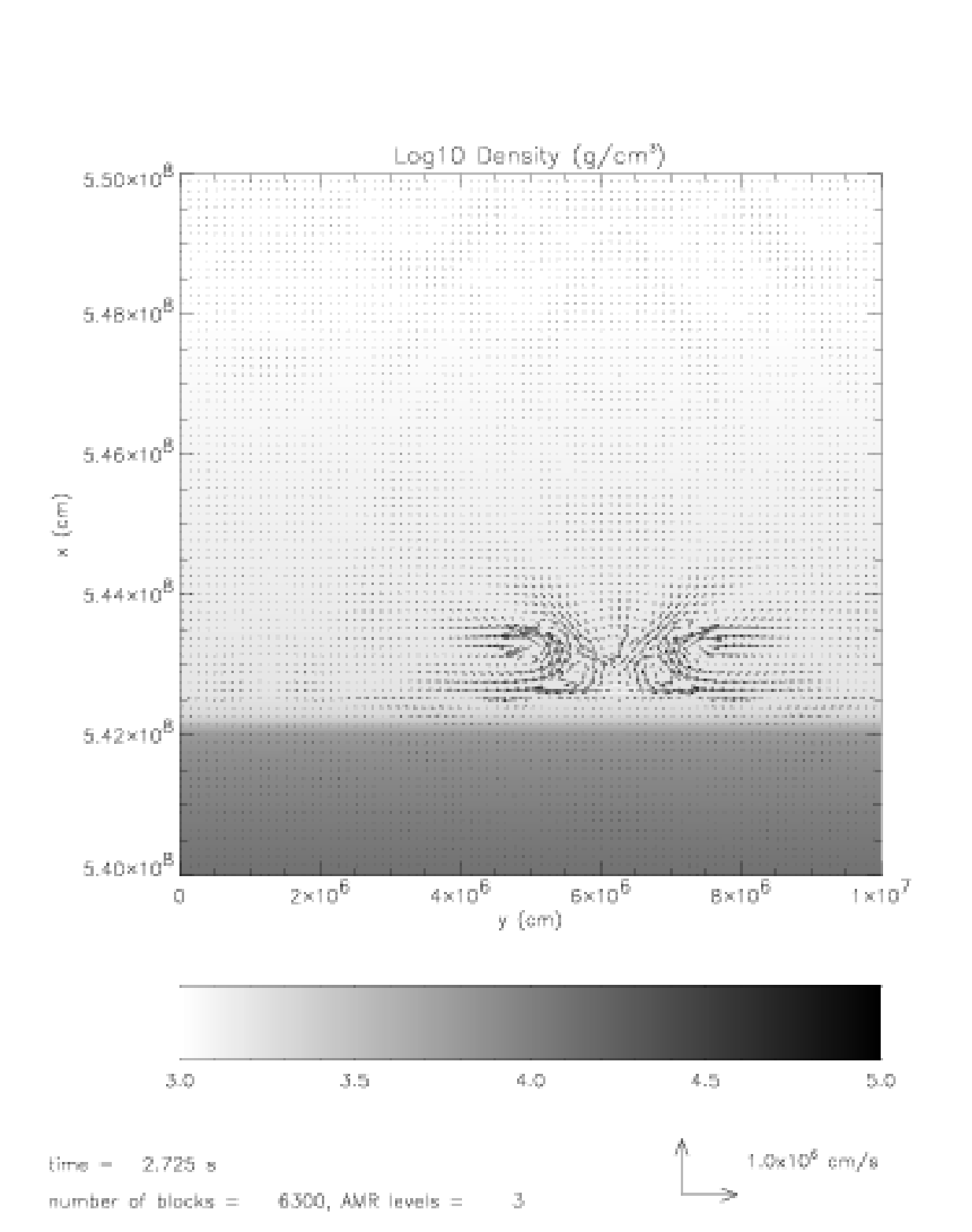}
  \includegraphics[width=0.32\textwidth]{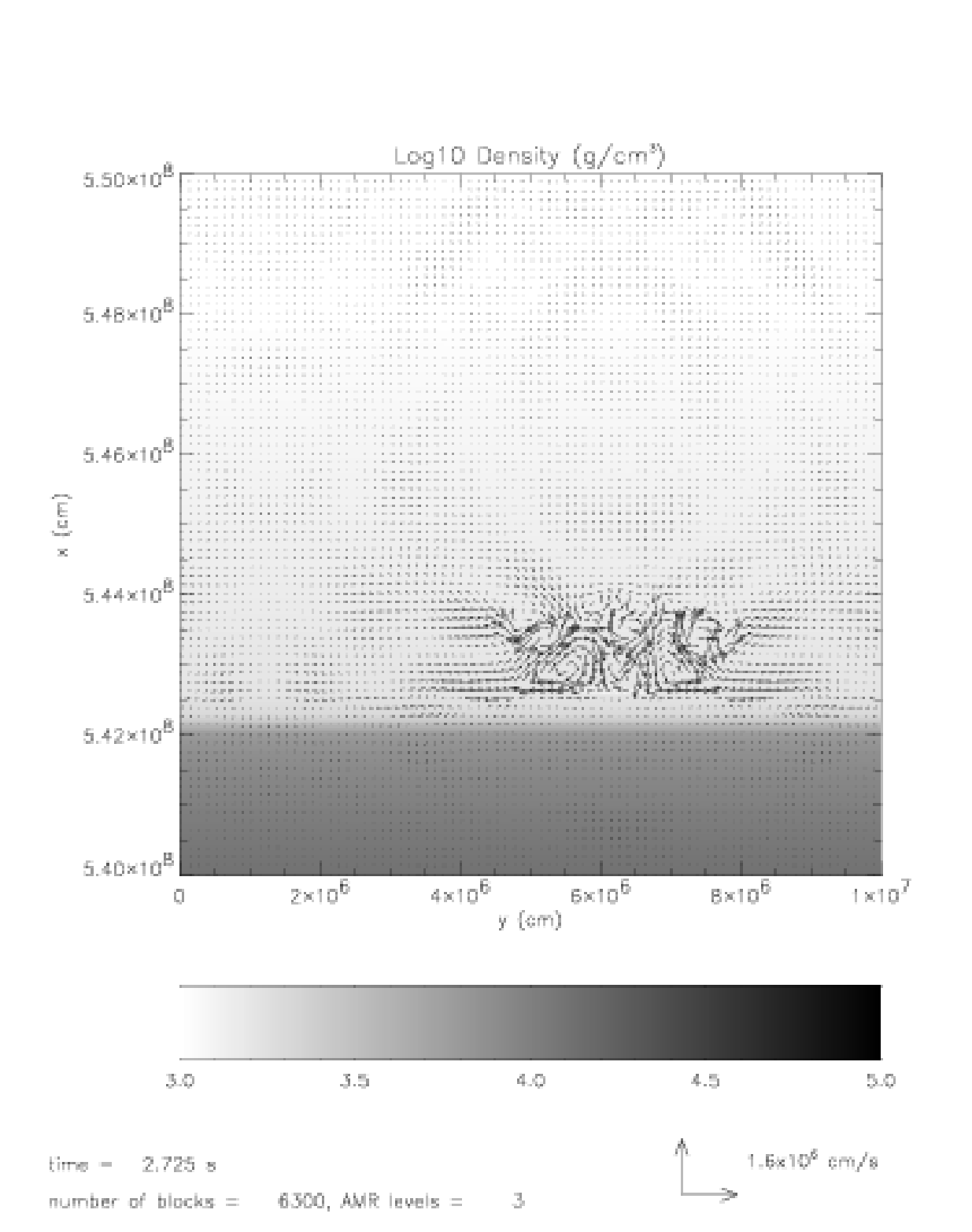}
  \includegraphics[width=0.32\textwidth]{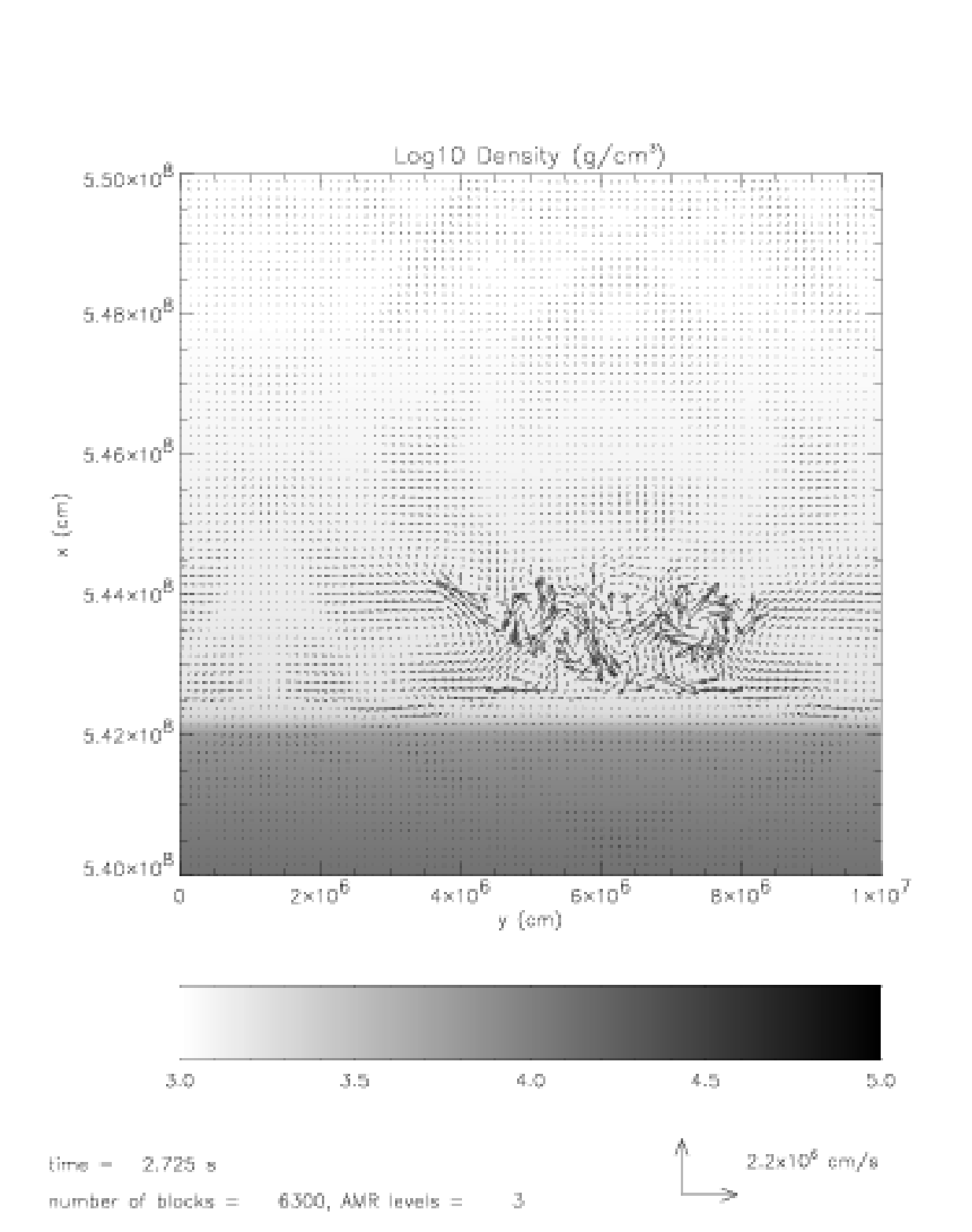}
  \caption{Early convective motions as in Figure \ref{fig:5percenttimeseries}
           at time $t = 1.2 {\mathrm{s}}$ after perturbation, for perturbations
           in the temperature of 2\%, 5\%, and 10\%.   Velocity arrows are
           scaled to the amount of thermal energy in the perturbation.}
  \label{fig:varyamplitude} 
\end{figure}

The dynamics of the rolls depend on the initial amplitude of the
perturbation.  Shown in Figure \ref{fig:varyamplitude} are the motions
in the atmosphere at time $t = 1.2~\mathrm{s}$ with temperature perturbations of
2\%, 5\%, and 10\%, respectively.  The velocity vectors in the plots
are scaled so that the kinetic energy of the motions are scaled to the
thermal energy of the perturbation; note that the sound waves have very
similar amplitudes in these plots.  Even with these scalings, the 10\%
perturbation generates considerably more motion.

\begin{figure}
  \includegraphics[width=0.35\textwidth]{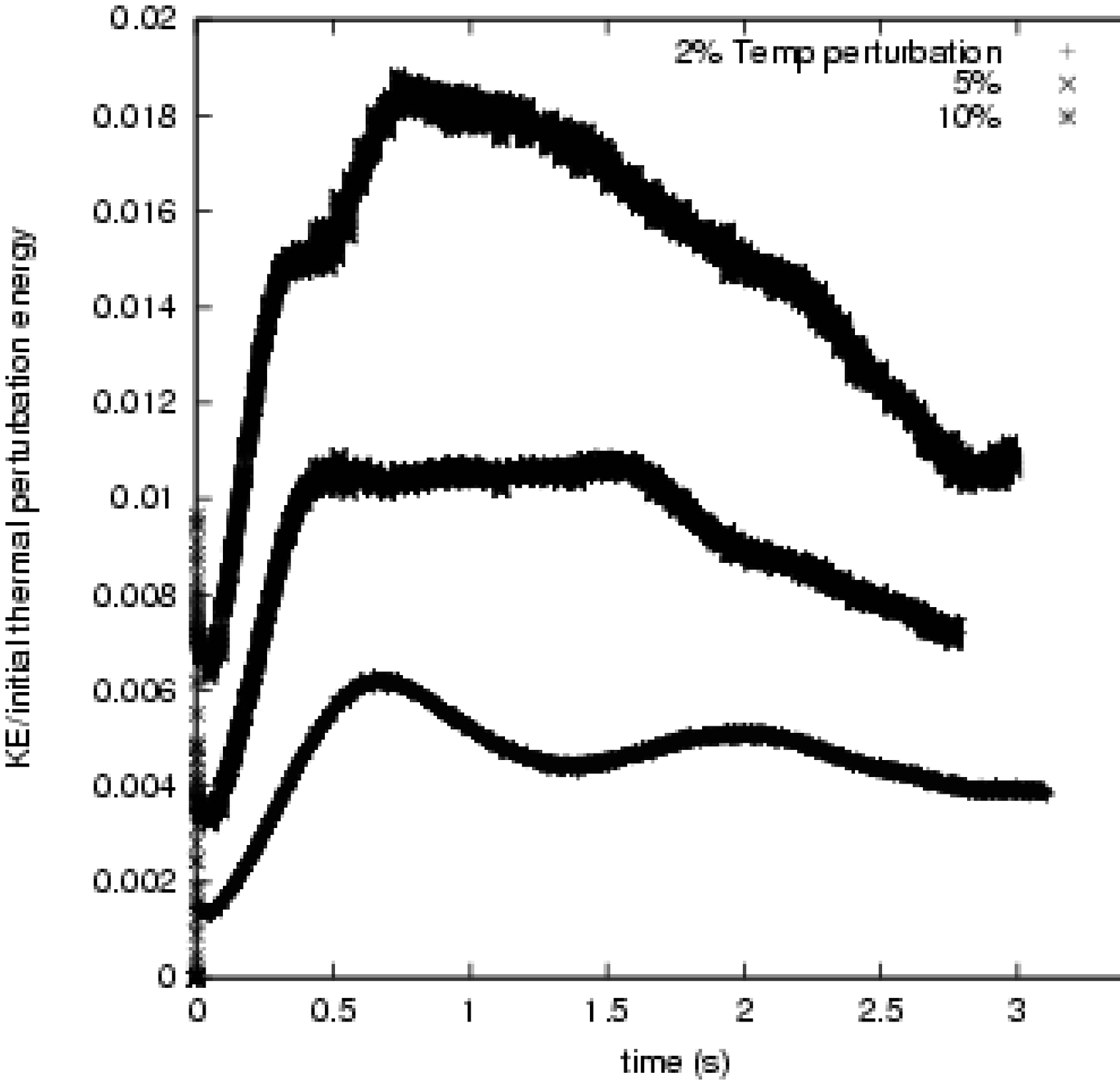} \includegraphics[width=0.45\textwidth]{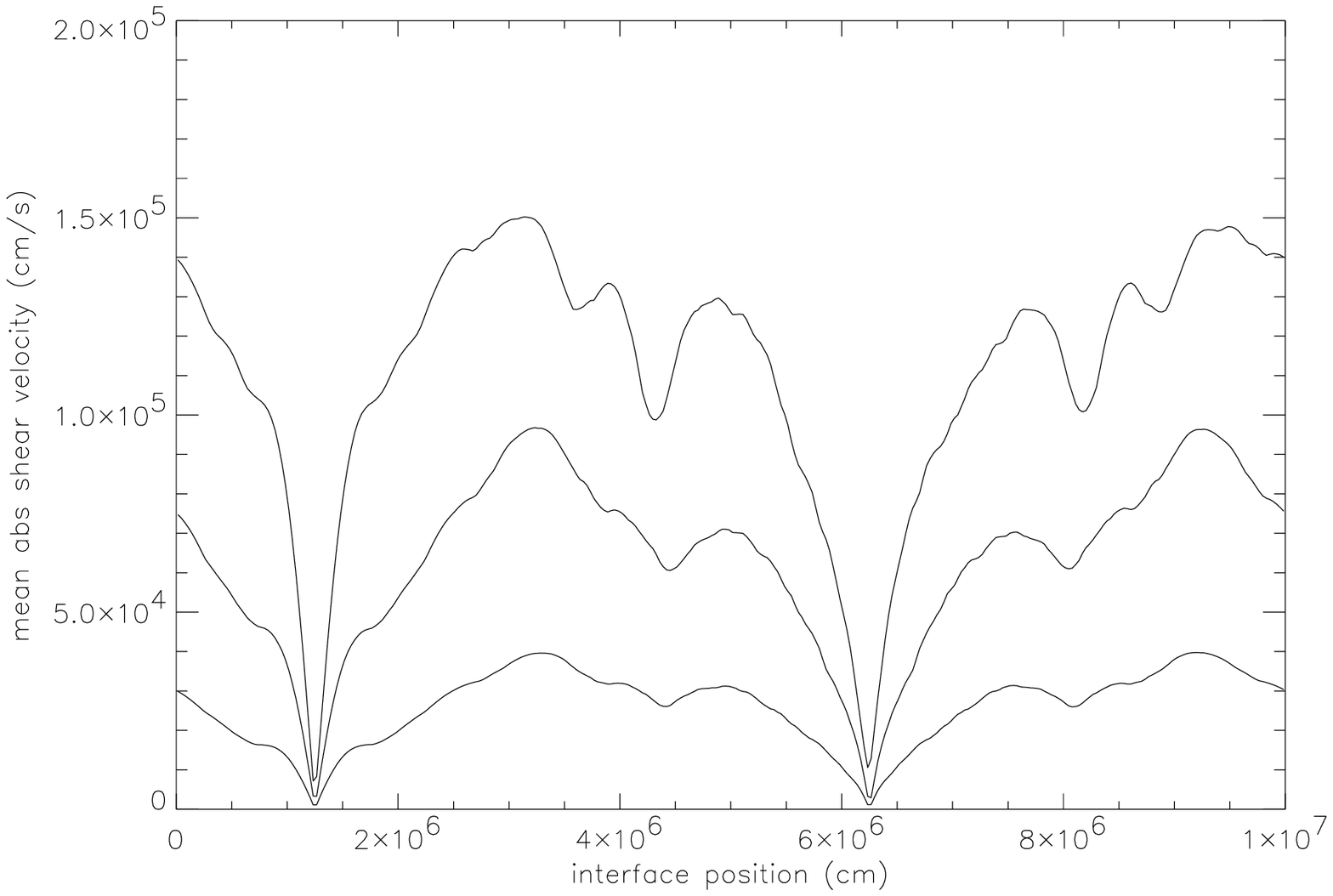}
  \caption{On the left, kinetic energy in the atmosphere measured in units
           of the original thermal perturbation energy for the three 
           perturbation amplitudes; on the right, the absolute values of the 
           interface velocities at time $t = 1.2 {\mathrm{s}}$ for the
           three amplitudes.   Large perturbations more efficiently cause
           motions, and the interfacial velocities rise monotonically with
           the perturbation size.}
  \label{fig:kevelocities} 
\end{figure}

We can quantitatively see the effects of perturbation size on convective
motions above.  Shown, in Figure \ref{fig:kevelocities} are the
kinetic energy as a fraction of the initial thermal energy perturbation,
and the average of the interfacial shear velocities over the course of
the simulation.   Larger perturbations produce motions at the interface
much more efficiently.

\section{Direction of Future Work}

\begin{figure}[ht]
  \includegraphics[width=0.6\textwidth]{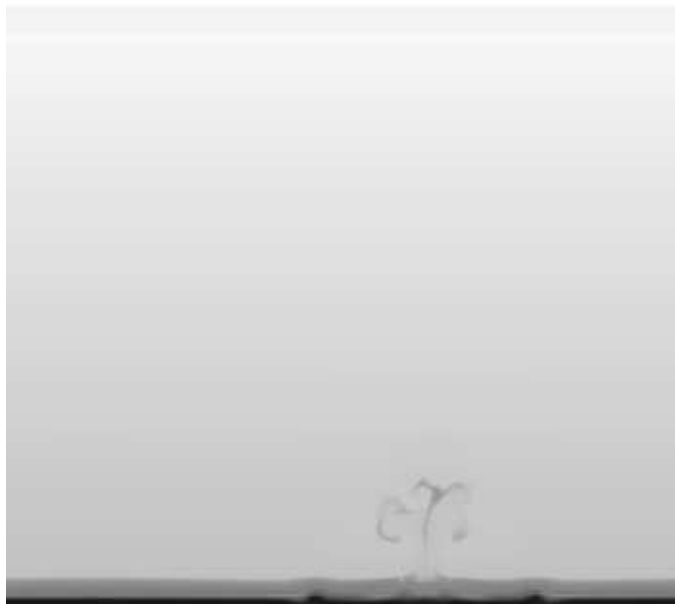}
  \caption{A metallicity plot at time $t = 0.89 {\mathrm {s}}$ of a 
           simulation similar to that shown in Figure 
           \ref{fig:5percenttimeseries}, but with the C+O white dwarf
           removed from the simulation domain; all metallicity is either
           from the accreted material, or from the preliminarly sub-grid
           model which represents the results of mixing simulations 
           presented elsewhere in this volume.}
  \label{fig:metallicity}
\end{figure}

Using what we are learning about shear and gravity-wave driven mixing
\cite{gravwavebreaking},
we can hope to model the unresolved mixing due to the interfacial shear
generated by these motions.  Shown in Figure \ref{fig:metallicity} is a
logarithmic plot of metallicity in the atmosphere at time 
$t = 0.89~\mathrm{s}$, where
the only source of C+O in the simulation is modelled mixing driven by the
velocities at the interface between the white dwarf and the atmosphere.
The sub-grid model we have used here is preliminary, based on early
results of observed mixing fluxes in small-scale simulations run
by A. Alexakis and A. C. Calder described in this volume.  As our
understanding of the convective motions and the shear-driven mixing
improves, we hope to see if this provides a robust dredge-up mechanism.
We will incorporate our improved subgrid model into both one-dimensional
and multi-dimensinoal simulations, and test our model results with respect
to the typical observed levels of enrichment of nova envelopes.

\subsection*{Acknowledgments}

This work is supported in part by the U.S. Department of Energy under
Grant No. B341495 to the Center for Astrophysical Thermonuclear Flashes at
the University of Chicago.  LJD is supported by the Krell Institute CSGF.
K. Olson acknowledges partial support from NASA grant NAS5-28524.
We greatfully acknowledge Ami Glasner for the initial model on which these
calculations were based.

\bibliographystyle{aipproc}

\bibliography{nova}


\end{document}